Review

# Natural Language Processing of Clinical Notes on Chronic Diseases: Systematic Review

Seyedmostafa Sheikhalishahi[1,2], MSc; Riccardo Miotto[3], PhD; Joel T Dudley[3], PhD; Alberto Lavelli[4], MSc; Fabio Rinaldi[5], PhD; Venet Osmani[1], PhD

[1]eHealth Research Group, Fondazione Bruno Kessler Research Institute, Trento, Italy
[2]Department of Information Engineering and Computer Science, University of Trento, Trento, Italy
[3]Institute for Next Generation Healthcare, Department of Genetics and Genomic Sciences, Icahn School of Medicine at Mount Sinai, New York, NY, United States
[4]NLP Research Group, Fondazione Bruno Kessler Research Institute, Trento, Italy
[5]Institute of Computational Linguistics, University of Zurich, Zurich, Switzerland

**Corresponding Author:**
Venet Osmani, PhD
eHealth Research Group
Fondazione Bruno Kessler Research Institute
Via Sommarive 18, Povo
Trento, 38123
Italy
Phone: 39 0461 31 2479
Email: vosmani@fbk.eu

## Abstract

**Background:** Novel approaches that complement and go beyond evidence-based medicine are required in the domain of chronic diseases, given the growing incidence of such conditions on the worldwide population. A promising avenue is the secondary use of electronic health records (EHRs), where patient data are analyzed to conduct clinical and translational research. Methods based on machine learning to process EHRs are resulting in improved understanding of patient clinical trajectories and chronic disease risk prediction, creating a unique opportunity to derive previously unknown clinical insights. However, a wealth of clinical histories remains locked behind clinical narratives in free-form text. Consequently, unlocking the full potential of EHR data is contingent on the development of natural language processing (NLP) methods to automatically transform clinical text into structured clinical data that can guide clinical decisions and potentially delay or prevent disease onset.

**Objective:** The goal of the research was to provide a comprehensive overview of the development and uptake of NLP methods applied to free-text clinical notes related to chronic diseases, including the investigation of challenges faced by NLP methodologies in understanding clinical narratives.

**Methods:** Preferred Reporting Items for Systematic Reviews and Meta-Analyses (PRISMA) guidelines were followed and searches were conducted in 5 databases using "clinical notes," "natural language processing," and "chronic disease" and their variations as keywords to maximize coverage of the articles.

**Results:** Of the 2652 articles considered, 106 met the inclusion criteria. Review of the included papers resulted in identification of 43 chronic diseases, which were then further classified into 10 disease categories using the *International Classification of Diseases, 10th Revision*. The majority of studies focused on diseases of the circulatory system (n=38) while endocrine and metabolic diseases were fewest (n=14). This was due to the structure of clinical records related to metabolic diseases, which typically contain much more structured data, compared with medical records for diseases of the circulatory system, which focus more on unstructured data and consequently have seen a stronger focus of NLP. The review has shown that there is a significant increase in the use of machine learning methods compared to rule-based approaches; however, deep learning methods remain emergent (n=3). Consequently, the majority of works focus on classification of disease phenotype with only a handful of papers addressing extraction of comorbidities from the free text or integration of clinical notes with structured data. There is a notable use of relatively simple methods, such as shallow classifiers (or combination with rule-based methods), due to the interpretability of predictions, which still represents a significant issue for more complex methods. Finally, scarcity of publicly available data may also have contributed to insufficient development of more advanced methods, such as extraction of word embeddings from clinical notes.

**Conclusions:** Efforts are still required to improve (1) progression of clinical NLP methods from extraction toward understanding; (2) recognition of relations among entities rather than entities in isolation; (3) temporal extraction to understand past, current, and future clinical events; (4) exploitation of alternative sources of clinical knowledge; and (5) availability of large-scale, de-identified clinical corpora.

**KEYWORDS**

electronic health records; clinical notes; chronic diseases; natural language processing; machine learning; deep learning; heart disease; stroke; cancer; diabetes; lung disease

## Introduction

### Overview

The burden of chronic diseases, such as cancers, diabetes, and hypertension, is widely accepted as one of the principal challenges of health care. While immense progress has been made in the discovery of new treatments and prevention strategies, this challenge not only persists, but its incidence is exhibiting an upward trend [1], with significant impact on patient quality of life and care costs. Consequently, there is a need for novel approaches to complement and go beyond current evidence-based medicine that can reduce the impact of chronic conditions on modern society.

A promising direction is the secondary use of electronic health records (EHRs) to analyze patient data, advance medical research, and better inform clinical decision making. Methods based in analysis of EHRs [2] are resulting in improved understanding of patient clinical trajectories [3] while enabling better patient stratification and risk prediction [4-6]. In particular, use of machine learning and especially deep learning to process EHRs is creating a unique opportunity to derive previously unknown clinical insights [7]. This is especially relevant for chronic diseases as their longitudinal nature provides a very large and continuous stream of data, where clinically meaningful patterns can be extracted and used to guide clinical decisions, including delaying or preventing disease onset.

However, EHRs are challenging to represent and model due to their high dimensionality, noise, heterogeneity, sparseness, incompleteness, random errors, and systematic biases. Moreover, a wealth of information about patient clinical history is generally locked behind free-text clinical narratives [8] since writing text remains the most natural and expressive method to document clinical events. Development of natural language processing (NLP) methods is essential to automatically transform clinical text into structured clinical data that can be directly processed using machine learning algorithms. Use of NLP in the clinical domain is seeing an increasing uptake with diverse applications, including identification of biomedical concepts from radiology reports [9], nursing documentation [10], and discharge summaries [11]. Frameworks based on NLP applied to clinical narratives, however, have not been widely used in clinical settings to help decision support systems or workflows.

### Motivation

Clinically relevant information from clinical notes has been historically extracted via manual review by clinical experts, leading to scalability and cost issues. This is of particular relevance for chronic diseases since clinical notes dominate over structured data (for example, Wei et al [12] graphically quantify the amount of clinical notes over structured data for chronic diseases such as rheumatoid arthritis, Parkinson disease, and Alzheimer disease). Availability of these data creates an immense opportunity for NLP to automatically extract clinically meaningful information that may delay or prevent disease onset, giving rise, however, to several challenges. In this paper we aimed to identify directions that could speed up the adoption of NLP of clinical notes for chronic diseases and provide an understanding of the current challenges and state of the art.

Systematic reviews related to processing of clinical notes have been published in the past [13-18]; however, none have focused specifically on chronic diseases, making it difficult to derive conclusions and recommendations on this specific and very diverse domain. In particular, this paper investigates NLP challenges related to 43 unique chronic diseases identified by our systematic review and discusses the trends of applying various NLP methods for clinical translational research. Based on the outcomes of this review, we also devised a number of recommendations on future research directions, including (1) evolution of clinical NLP methods from extraction toward understanding; (2) recognition of relations among entities, rather than entities in isolation; (3) temporal extraction in order to understand past, current, and future clinical events; (4) exploitation of alternative sources of clinical knowledge; and (5) availability of large-scale deidentified and annotated clinical corpora.

## Methods

### Search Strategy and Information Sources

We followed the Preferred Reporting Items for Systematic Reviews and Meta-Analyses (PRISMA) guidelines [19]. We carried out a search of several databases to identify all potentially relevant articles published from January 1, 2007, to February 6, 2018, including Scopus, Web of Science (including MEDLINE) and PubMed, and the Association for Computing Machinery (ACM) Digital Library. We have limited the search to journal articles written in English. In all the searches we used the combination of the following groups of keywords: (1) "clinical notes," "medical notes," or "clinical narratives"; (2) "natural language processing," "medical language processing," "text mining," or "information extraction"; and (3) "chronic disease," "heart disease," "stroke," "cancer," "diabetes," or "lung disease" (where the last set of keywords reflects the top five chronic diseases). The search keywords were selected to be exhaustive to maximize coverage of the articles. The exact queries are provided in Multimedia Appendix 1.

**Article Selection**

In the initial queries we also included the following terms: "electronic health records," "EHR," "electronic medical records," and "EMR." This led to a total of 2652 retrieved articles. However, upon reviewing these articles, we noticed that the scope was too broad, providing results outside of focus of this review. Consequently, we narrowed the search strategy to the keywords specified in the previous section, obtaining a total of 478 articles, with 401 articles from Scopus, 58 from Web of Science (including PubMed), 13 from ACM Digital Library, and 6 added manually, including 4 conference papers.

After removing 46 duplicates, 432 articles were retained, and two authors (MS and VO) reviewed their titles and abstracts (216 articles each). After this screening phase, 159 articles were retained for further analysis.

In the second screening stage, five authors independently reviewed the 159 full-text articles, resulting in 106 articles fulfilling our criteria that are discussed in this review. The most common reason for exclusion was that the work was not directly related to chronic diseases (n=32); another reason was the work was not topical (eg, the article was not a journal paper or we could not retrieve the text). A flowchart and description of the selection process are provided in Figure 1 and Multimedia Appendix 2, respectively.

**Figure 1.** Preferred Reporting Items for Systematic Reviews and Meta-Analyses article selection flowchart. ACM: Association for Computing Machinery; NLP: natural language processing.

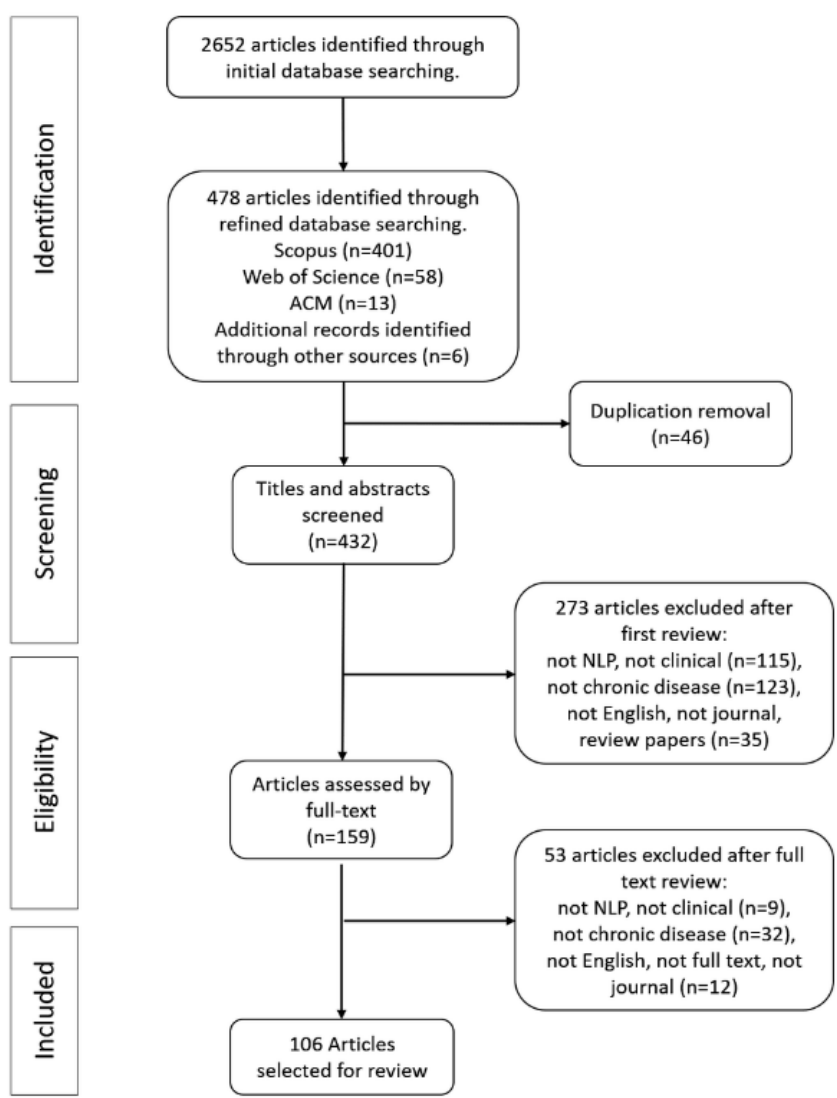

## Results

### Categorization of Diseases

The 106 articles reviewed were largely related to 43 unique chronic diseases (as shown in Multimedia Appendix 2). One of our aims was to understand the extent of NLP for specific disease categories and their associated clinical notes. Therefore, we grouped the 43 unique chronic diseases into 10 disease categories using the *International Classification of Diseases, 10th Revision* (ICD-10) as shown in Table 1.

**Table 1.** Classifications of chronic conditions studied (n=102) and the corresponding number of papers found.

| Classification of chronic condition | Studies, n (%) | Conditions included |
| --- | --- | --- |
| Diseases of the circulatory system | 38 (35.8) | Congestive heart disease (2), coronary artery disease (6), heart disease (6), heart failure (7), hypertension (5), peripheral arterial disease (3), pulmonary disease (4) |
| Neoplasms | 34 (32.1) | Breast cancer (8), colorectal cancer (7), prostate cancer (4), lymphoma (2) |
| Endocrine, nutritional, and metabolic diseases | 14 (13.2) | Type 2 diabetes mellitus (12), obesity (2) |
| Other diseases | 16 (15.1) | Diseases of the digestive system (3), diseases of the genitourinary system (3), diseases of the musculoskeletal system and connective tissue (3), diseases of the respiratory system (2), mental and behavioral disorders (2), multidisease (3) |

**Figure 2.** Relationship between chronic diseases (black sectors) and articles included in the review (for clarity we have included only diseases that are addressed by three or more articles).

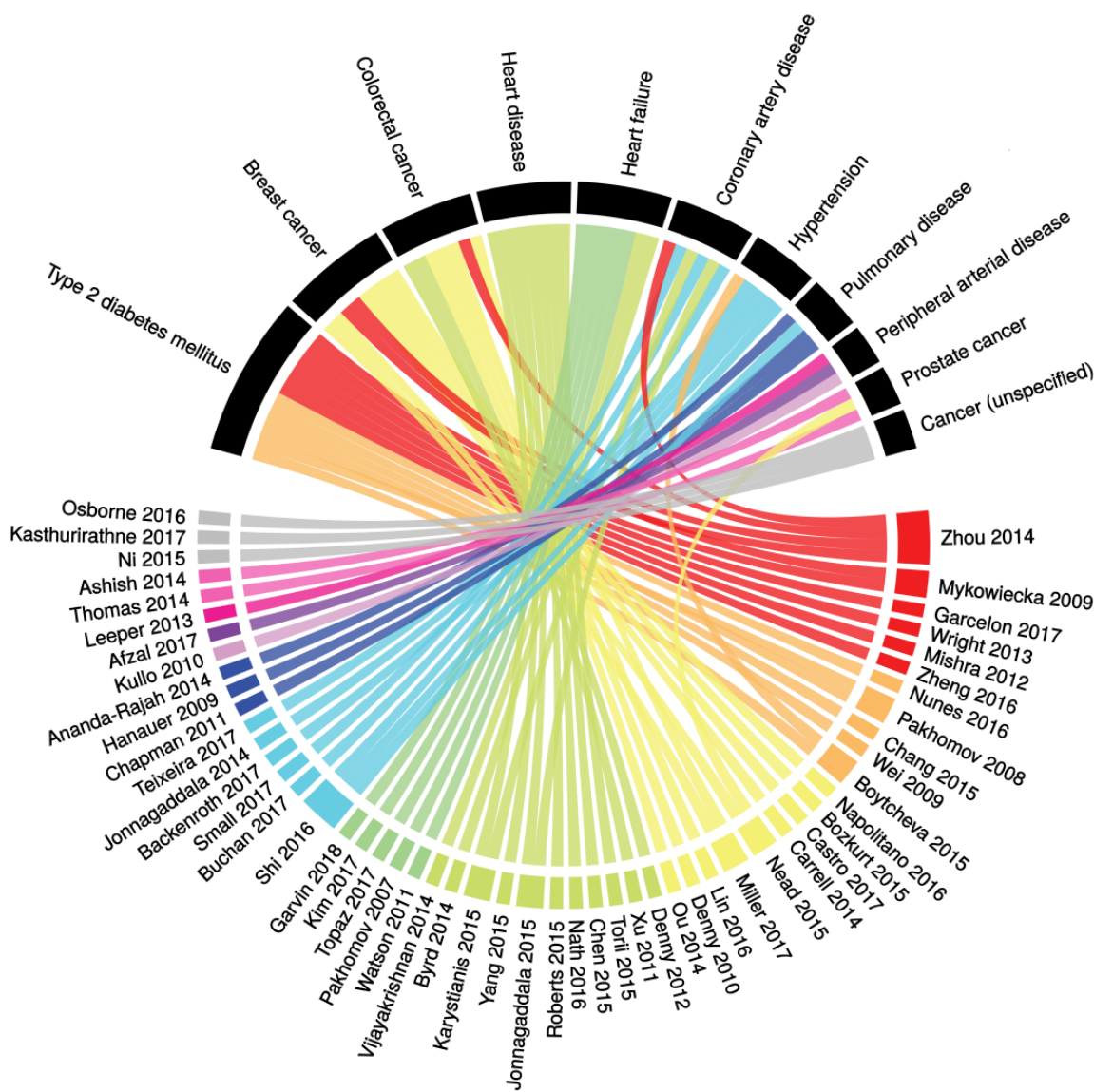

The top three disease groups were (1) diseases of the circulatory system (n=38) (such as coronary artery disease [20] and hypertension [21]); (2) neoplasms (n=34) (such as breast cancer [22] and prostate cancer [23]); and (3) endocrine, nutritional, and metabolic diseases (n=14) (such as type 2 diabetes [24] and obesity [25]). An overview of the diseases studied and the corresponding articles is shown in Figure 2.

An unexpected finding is that despite the higher incidence of metabolic diseases in the general population [26] compared with diseases of circulatory system [27], the use of NLP in clinical narratives of these diseases exhibits an opposite trend. Diseases of the circulatory system are represented in much greater numbers with respect to metabolic diseases (n=38 vs n=14, respectively). We hypothesize that the structure of data contained in EHRs may explain this finding. Medical records related to metabolic diseases typically contain much more structured data (for example, numerical values for various physiological and physical parameters) than medical records for diseases of the circulatory system, which focus more on unstructured data [28]. This creates a more pressing need to use NLP to extract information from notes related to diseases of the circulatory system, whereas EHRs of patients with metabolic diseases in large part may already contain data that can be used by algorithms with minimal preprocessing. In the sections that follow we summarize the most representative papers (the complete list is provided in Multimedia Appendix 2).

## Disease Groups

### Diseases of the Circulatory System

#### Cardiovascular Diseases

Most of the work in this area focused on using NLP to estimate the risk of heart disease. As an example, Chen et al [29] developed a hybrid pipeline based on both machine learning and rules to identify medically relevant information related to heart disease risk and track the disease progression over sets of longitudinal patient records, including clinical notes (similarly to Torri et al [30]). Karystianis et al [31] and Yang et al [32] evaluated the identification of heart disease risk factors from the clinical notes of diabetic patients. In a slightly different approach, Roberts et al [33] focused on estimating heart disease risk based on classification of 8 risk triggers (for example, aspirin). Other studies in this area have focused on evaluating the use of aspirin as a risk factor [34,35], extracting heart function measurements from echocardiograms [36], deep vein thrombosis and pulmonary embolism [37], and low-density lipoprotein level and statins use [38].

Risk of stroke and major bleeding in patients with atrial fibrillation has been predicted using structured data and clinical notes [39], while patients with heart failure have been identified using clinical notes only [40]. Moreover, medical reports written in the Italian language have been used to identify arrhythmia events [41].

#### Peripheral and Coronary Arterial Disease

Several studies used NLP to extract cases of peripheral arterial disease (PAD) and critical limb ischemia from clinical notes [42,43], including a genome-wide associated study, focusing on PAD to identify drugs, diseases, signs/symptoms, anatomical sites, and procedures [44]. Leeper et al [45] used NLP to identify PAD patients to conduct a safety surveillance study on exposure to Cilostazol, finding complications of malignant arrhythmia and sudden death not observed in association with the drug. Furthermore, Clinical Text Analysis Knowledge Extraction System (cTAKES) has been used to process clinical history of diabetic patients to predict development of PAD [46].

#### Hypertension

Work on hypertension has been principally focused on NLP to extract relevant indicators, comorbidities, and drug therapies [21]. Analysis of clinical narratives in the Bulgarian language of 100 million outpatient notes was used to extract numerical blood pressure values with a high sensitivity and recall [47], while term hypertension was extracted from free-text notes, using a rule-based, open-source tool [48]. Clinical notes and several types of medical documents were also used to identify hypertensive individuals using open-source medication information extraction (IE) system MedEx [49].

#### Right-Sided, Left-Sided, and Congestive Heart Failure

Byrd et al [50] and Jonnagaddala et al [20] proposed a hybrid NLP model to identify Framingham heart failure signs and symptoms from clinical notes and EHRs (ie, classifying whether Framingham criteria are asserted). Left ventricular ejection fraction was extracted from free-text echocardiogram reports [51], while unstructured, longitudinal EHRs of diabetic patients were used to extract relevant information of heart disease, using naïve Bayes and conditional random field (CRF) classifiers [52].

Wang et al [53] proposed a system for the identification of congestive heart failure (CHF) from EHRs, which they prospectively validated. Furthermore, left ventricular ejection fraction plus the associated qualitative and quantitative values were used to identify patients at risk of CHF [54], while free-text notes were used to distinguish left and right heart failure [55].

#### Heart Failure Identification

Topaz et al [56] developed an algorithm to identify heart failure (HF) patients with ineffective self-management of diet, physical activity, adherence to medication, and clinical appointments using discharge summary notes, while Garvin et al [57] focused on the quality of care for HF patients. Vijayakrishnan et al [58] explored the application of a previously validated text and data-mining tool to identify the presence of HF signs and symptoms criteria in the EHRs of a large primary care population. They found that HF signs and symptoms were documented much more frequently among the eventual HF cases, years before the first diagnosis as well, thus suggesting a potential future role for early detection of HF. Last, regular expressions were used to identify predefined psychosocial factors that served as predictors of the likelihood to be readmitted to the hospital after a case of HF [59].

### Neoplasms

#### Overview

This section reviews a number of cancer-related studies, including detection of multiple types of cancer [60,61], extracting tumor characteristics and tumor-related information [62-64], disease trajectories of patients with cancer [65], cancer recurrence [23,66], and detection of stage of cancer [67,68].

Kasthurirathne et al [60] evaluated the performance of common classification algorithms to detect cancer cases from free-text pathology reports using nondictionary approaches. Yim et al [62] explored a machine learning algorithm to extract tumor characteristics by applying reference resolution on radiology reports. Jensen et al [65] developed a methodology that allows disease trajectories of cancer patients to be estimated from the clinical text. Napolitano et al [67] facilitated the extraction of information relevant to cancer staging, proposing a model for semistructured reports that outperformed the model for unstructured reports alone.

A number of studies have focused on different applications of NLP in pathology, histopathology, and radiology reports [69], including extracting relevant domain entities from narrative cancer pathology reports [70], negation detection of medical entities in pathology reports [71], sentence translation from pathology reports into graph representations [72], extracting information from pathology reports and pathology classifications [73,74], and named entity recognition from histopathology notes [75].

The three most common types of cancers found are breast cancer (n=8), colorectal cancer (n=7), and prostate cancer (n=4).

**Breast Cancer**

Carrell et al [66] proposed an NLP system to process clinical text to identify breast cancer recurrences, while Castro et al [22] addressed the automated Breast Imaging-Reporting and Data System (BI-RADS) categories extraction from breast radiology reports. Miller et al [76] proposed a tool for coreference resolution in clinical texts evaluated within the domain (colon cancer) and between domains (breast cancer). Mykowiecka et al [77] propose a rule-based IE system evaluated on mammography reports. Bozkurt et al [78] developed NLP methods to recognize lesions in free-text mammography reports and extract their corresponding relationships, producing a complete information frame for each lesion.

**Colorectal and Prostate Cancer**

EHRs and NLP were used to identify patients in need of colorectal cancer screening [79] and detect colonoscopy-related concepts as well as temporal-related information [80]. Additionally, EHRs and NLP were used to also identify patients with prostate biopsies positive for prostatic adenocarcinoma [81].

**Liver and Pancreatic Cancer**

Ping et al [82] extracted textual information concerning a set of predefined clinical concepts from a variety of clinical reports for patients with liver cancer, while Al-Haddad et al [83] identified patients with confirmed surgical pathology diagnoses of intraductal papillary mucinous neoplasms.

*Endocrine, Nutritional, and Metabolic Diseases*

Applications of NLP in the domain of endocrine, nutritional, and metabolic diseases include negation detection and mention of family history in free-text notes [84] and assigning temporal tags to medical concepts [85]; obesity [25,86] and diabetes identification [77,87-89]; and diabetes complications such as foot examination findings [90], vision loss [91], and quantifying the occurrence of hypoglycemia [24].

Two support vector machines (SVMs) were combined to automatically identify obesity types by extracting obesity and diabetes-related concepts from clinical text [86] in addition to patient identification [92]. An SVM-based system was developed and validated to identify EHR progress notes pertaining to diabetes [87], while foot examination findings from clinical reports [90] were used to predict quality of life [93]. Additionally, an analysis of a large EHR database was used to quantify occurrence of hypoglycemia [24].

*Other Disease Categories*

The remaining 16 papers focused on processing clinical notes of different types of chronic diseases. Three studies concern diseases of the musculoskeletal system and connective tissue, in particular classification of snippets of text related to axial spondyloarthritis in the EMRs of US military veterans using NLP and SVM [94], phenotyping systemic lupus erythematosus [95], and identification of rheumatoid arthritis patients via ontology-based NLP and logistic regression [96]. In the domain of diseases of the digestive system, Chen et al [97] used natural language features from pathology reports to identify celiac disease patients, Soguero-Ruiz et al [98] used feature selection and SVMs to detect early complications after colorectal cancer, and Chang et al [99] integrated rule-based NLP on notes with ICD-9s and lab values in an algorithm to better define and risk-stratify patients with cirrhosis.

Two papers evaluated deep learning in a multidisease domain. In particular, Miotto et al [3] derived a general purpose patient representation from aggregated EHRs (structured clinical data and clinical notes) based on neural networks that facilitates clinical predictive modeling given the patient status. Clinical notes were parsed using the National Center for Biomedical Ontology's Open Biomedical Annotator to extract medical terms and further processed using topic modeling (latent Dirichlet allocation). Shi et al [100] proposed assessing disease risk from patient clinical notes using word embeddings and convolutional neural networks with full connection layer.

Neural networks were also used to process clinical notes for phenotyping psychiatric diagnosis [101]. In particular, this model included two neural networks, one highly accurate at rejecting patients but poor at identifying suitable ones and the other one with the opposite capabilities. In the same domain of mental and behavioral disorders, comorbidity networks were derived from the patient notes at the largest Danish psychiatric hospital in order to extract disease correlations [102].

IE from clinical notes based on NLP was also used to (1) screen computed tomography reports for invasive pulmonary mold [103], (2) discover the co-occurrences of chronic obstructive pulmonary disease with other medical terms [104], (3) quantify the relationship between aggregated preoperative risk factors and cataract surgery complications [105], (4) detect patients with multiple sclerosis from the clinical notes prior to the initial recognition by their health care providers [106], and (5) identify patients on dialysis in the Multiparameter Intelligent Monitoring in Intensive Care II (MIMIC-II) publicly available dataset [107].

Last, Pivovarov and Elhadad [108] used clinical notes of patients with chronic kidney disease to validate a novel model to compute the similarity of two medical concepts by combining complementary information derived from usage patterns of clinical documentation, accepted definitions, and position of the concepts in an ontology.

**Information Extraction Methods**

In order to understand trends in NLP methods for chronic diseases, in this review we have analyzed papers with respect to the methods employed (machine vs rule-based learning). While there is an increasing use of machine learning methods in comparison to rule-based (as shown in Figure 3), it is not as pronounced as we had expected considering the superior performance of machine learning algorithms shown in the NLP literature [109]. This result may reflect the fact that we are still currently witnessing a transition from rule-based methods to machine learning algorithms, with rule-based methods used as a baseline to compare the performance of machine learning approaches.

Our review identified 16 papers that employed hybrid approaches combining rule-based and machine learning methods. Out of these, 2 papers describe work to identify diseases, risk factors, medications, and time attributes. In particular, a hybrid pipeline based on CRFs, SVMs, and rule-based approaches was used to identify negation information and normalize temporal expressions [29], while a series of SVM models in conjunction with manually built lexicons were used to classify triggers specific to each risk factor [33].

We identified 24 papers that focused on comparison between performance of rule-based and machine learning methods. Typically, the rule-based methods were used as a baseline to test the performance against machine learning algorithms.

As for rule-based approaches, the methods in this review include dictionary lookup [110-112], terminology identification based on domain ontologies [3,42,45,58], various types of manually defined rules [37,113], and regular expressions patterns [114,115].

The most widely used machine learning approach is SVMs, having been used for predicting heart disease in medical records [32,46], identifying EHR progress notes pertaining to diabetes [94], and categorizing breast radiology reports according to BI-RADS [22].

Naïve Bayes was the second most frequent approach, being used to predict heart disease in medical records [30,80], classify smoking status [52], search EMR records to identify multiple sclerosis [106], and classify EMR records for obesity [86] and cancer [60,65,67]. CRFs are the third most frequent approach, have been used to predict heart disease in medical records [29,32], identify EHR progress notes pertaining to diabetes [85], categorize breast radiology reports [22], and identify tumor attributes in radiology reports [63]. Lastly, random forests were used for predicting heart disease [53], classifying cancer types [60], and identifying hypertension [49].

Figure 3. Natural language processing rule-based methods versus machine learning for chronic diseases.

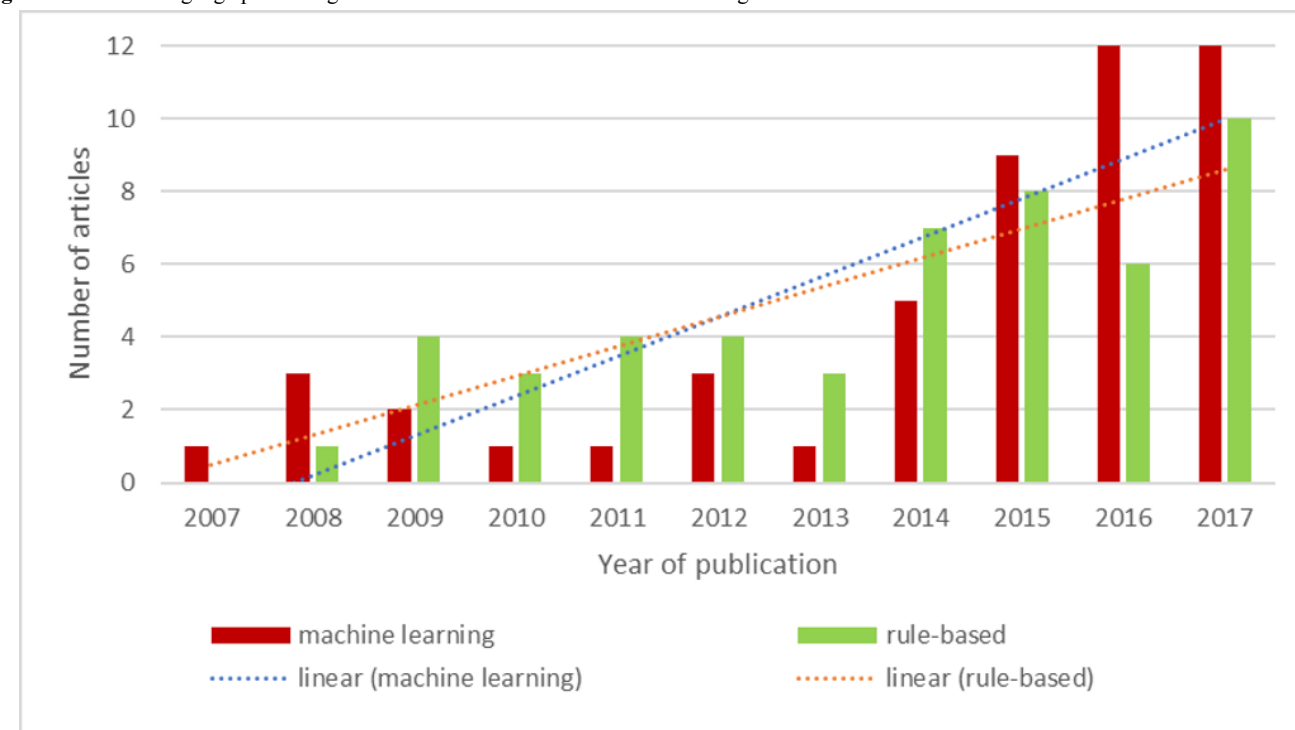

**Table 2.** Most frequently used natural language processing methods and the corresponding number of papers.

| Method | Papers (n) |
| --- | --- |
| Support vector machine | 18 |
| Naïve Bayes | 11 |
| Conditional random fields | 7 |
| Random forest | 4 |
| Maximum entropy | 3 |
| Decision tree | 3 |
| Deep neural networks | 3 |
| Logistic regression | 3 |
| Rule-based methods | 74 |

It is interesting to note that there are only 3 papers using approaches based on deep learning [3,100,101], as shown in Table 2. In particular, Geraci et al [101] apply deep neural networks to EMRs to identify suitable candidates for a study on youth depression. Miotto et al [3] present a method to derive a patient representation that facilitates clinical predictive modeling from aggregate EHRs, including clinical narratives. They represented free-text notes using topic modeling. This method significantly outperformed those achieved by standard feature learning strategies. Finally, Shi et al [100] propose a disease assessment model based on clinical notes, using convolutional neural network for disease risk assessment. The experiment involved patients with cerebral infarction, pulmonary infection, and coronary atherosclerotic heart disease.

### Natural Language Processing Tasks, Methods, and Datasets

The NLP works described in the reviewed papers and associated approaches reveal that the most frequently described tasks are text classification and entity recognition. The majority of the papers describe text classification tasks using standard approaches in NLP such as SVM (n=12) and naïve Bayes (n=4). Entity recognition approaches are based on manually developed resources (dictionary, regular expressions, handwritten rules) as well as methods based on machine learning. As for the former, there are dictionary-based approaches (n=5) and those relying on regular expressions (n=12). As for the latter, the approaches are mainly based on standard machine language techniques such as CRF and deep learning. A few papers describe approaches to coreference resolution (n=2) and negation detection (n=3). Coreference resolution is addressed using SVM, while negation detection is based on SVM (n=2) or manual rules (n=1).

Regarding datasets, the majority of the papers describe experiments run on datasets that are not publicly available (typically clinical data collected at research-based health care institutions and exploited by in-house NLP teams). On the other hand, out of 16 papers involving publicly available corpora, 12 exploit the Informatics for Integrating Biology and the Bedside (i2b2) datasets. The other 4 public datasets used are MIMIC-II [107], PhenoCHF [116], Temporal Histories of Your Medical Event (THYME), and Cancer Deep Phenotype Extraction (DeepPhe) [76].

### Comparisons to Other Systematic Reviews

Interest in using NLP for the automated processing of medical records, and in particular of free-text clinical notes, is increasing, exemplified by a number of recent reviews of the field. Yet none of these works focuses solely on chronic diseases, where the amount of patient clinical notes tends to be larger than other domains or provides specific recommendations on how to advance the field toward a clinical adoption that helps in treating people with chronic conditions. Here we briefly provide a summary of previous works partially related to the work presented in this paper.

Ford et al [13] present a systematic review of 67 papers using IE techniques applied to medical records for the purpose of case detection (ie, finding occurrences of specific medical conditions). Similarly, Kreimeyer et al [117] review 86 papers focusing on clinical NLP systems and a set of 71 associated NLP tasks.

The work by Shivade et al [14] reviews 97 papers aiming at identifying patient cohorts for further medical studies. Different from our work, theirs is not limited to investigation of studies using NLP and text mining but includes rule-based approaches, which do not make use of the textual part of the medical records. They observe, however, that the use of machine learning and statistical and NLP methods is on the rise compared to rule-based systems.

Abbe et al [118] consider applications of text mining in psychiatry through a PRISMA-based review. The study evaluates the application of specific NLP techniques in relation to the goal of the studies, first qualitatively, and then with a cluster analysis of the topics of selected abstracts. It identifies four main themes in the publications taken into consideration: (1) psychopathology (2), patient perspective, (3) medical records, and (4) medical literature. The scope of this review only partially overlaps with our own, given the narrow thematic analysis and inclusion of studies that deal with IE from other textual resources, such as patient perspectives.

The review by Spasic et al [119] focuses on cancer research. The authors classify the studies by cancer type and type of processed document. They do not focus solely on studies based on medical records or other types of clinical documents but also include meta-studies that apply text mining techniques to PubMed publications. They classify NLP applications in four categories: named entity recognition, IE, text classification, and information retrieval. Their investigation reveals a predominance of symbolic approaches (dictionary and rule-based).

The work by Pons et al [120] is a systematic review of NLP applications in the area of radiology. After initial preselection based on abstracts, a detailed review of the full text of the selected papers ultimately yields 67 publications, all deemed to consider practical applications of NLP in radiology. The selected publications are then grouped into five broad categories depending on the specific application: diagnostic surveillance, cohort building, query-case retrieval, quality assessment of radiological practice, and clinical support services. The authors provide a detailed comparative analysis of the performance reported in each publication, grouping them by application category.

Closest to our work is a systematic review by Wang et al [18] that has focused on IE applications; however, our review additionally includes methodologies used in analysis of clinical notes, providing a wider set of articles. We believe that our review has a broader and more recent coverage of chronic diseases, followed by detailed analysis for each disease, compared with previous reviews, which have focused on specific conditions such as cancer [119], psychiatry [118], radiology [120], or IE applications [18].

**Publication Venues**

The 106 articles considered in this review were published in 50 unique venues. Figure 4 illustrates how we manually sorted publication venues into three categories: (1) clinical medicine, (2) medical informatics, and (3) computer science. We observed that most of the studies were published in medical informatics journals. Figure 5 shows an increasing trend in number of publications over the years (except for the year 2018 due to partial-year retrieval) implying an increasing interest in the application of NLP in both clinical and informatics research for chronic diseases.

**Figure 4.** Categorization of the publication venues.

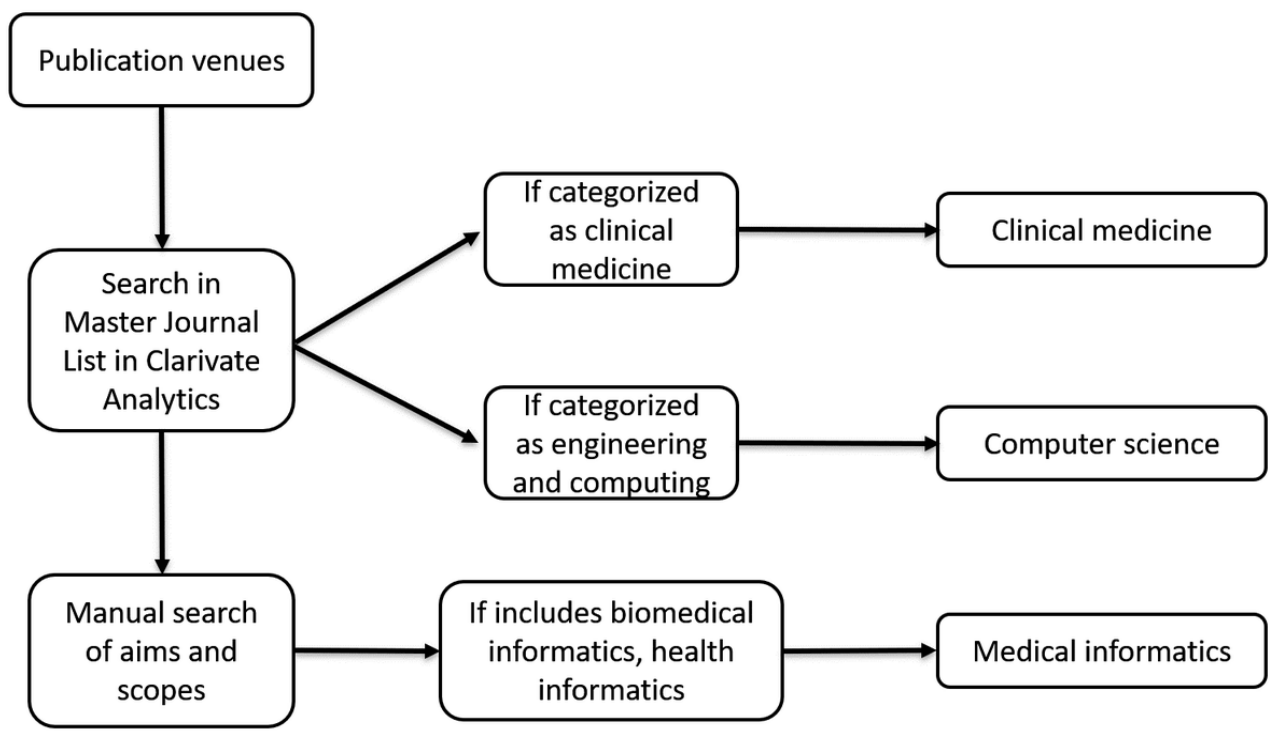

**Figure 5.** Distribution of included studies according to publication venues.

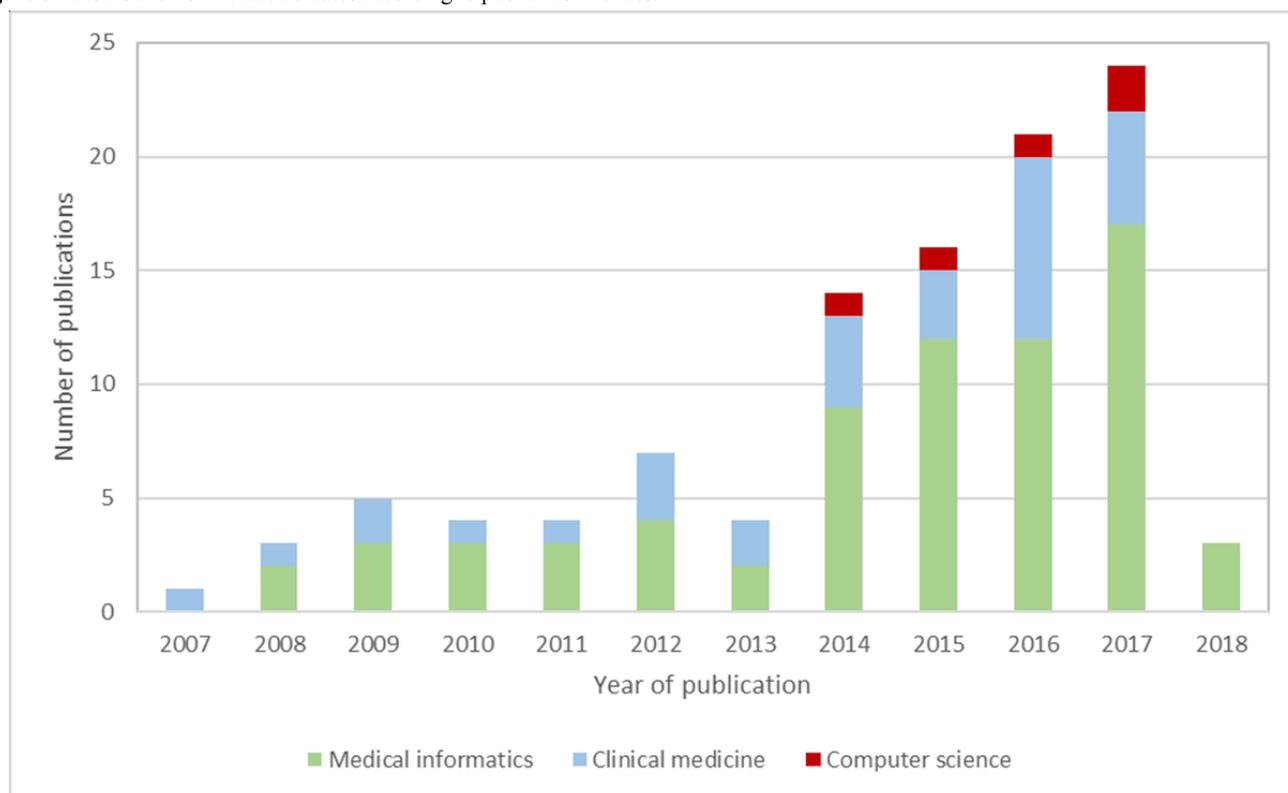

## Discussion

### Principal Findings

Our systematic review has shown that NLP has a wide range of applications for processing clinical notes of diverse chronic diseases (43 unique chronic diseases identified in the analysis). In this respect, there is a significant increase in the use of machine learning compared with rule-based methods. Despite the potential offered by deep learning, the majority of papers still rely on shallow classifiers. In fact, only a handful of studies (ie, 3 papers) made use of deep classifiers or general deep learning methods for NLP. This was unexpected, considering the potential of deep learning for text processing [121]. Our hypothesis is that since deep learning is still an emerging area, initial applications in the clinical domain may have been published in workshops, conference proceedings, and the e-print repository arXiv rather than journals, the focus of this review. In this respect, a keyword search in arXiv for "deep learning," and "clinical notes," "medical notes," or "clinical narratives" for the previous five years (2013-2018) shows a significant growth of papers: 7 from 2013 to 2015, 13 in 2016, 19 in 2017, and 22 in 2018. In addition, the longer review time for journals has likely contributed to this outcome for the more recent papers. We expect this result to shift in the coming years as an increasing amount of work based on deep learning to process clinical notes is published in peer-reviewed journals.

Another finding from our review is that the majority of papers reviewed identify risk factors for a particular disease and classify a clinical note by a certain disease phenotype. However, there are only a handful of papers that extract comorbidities from the free-text or integrate clinical notes with structured data for prediction and longitudinal modeling of trajectories of patients with chronic diseases. Such an outcome could be related to the use of data analysis methods and algorithms (such as shallow classifiers and rule-based approaches highlighted earlier) that do not have the capability to capture temporal and longitudinal relationships between clinical variables and in turn capture disease evolution. Tools (such as MetaMap) and methods (such as mapping n-grams to ontologies) used may have been other influencing factors. While these tools allow extracting meaningful medical information from the text, inherently they reduce the possibility to derive more complex relationships, principally due to phrase structure (for example "breast and lung cancer" may be identified only as "breast" and "lung cancer" rather than both "breast cancer" and "lung cancer"). However, the use of relatively simple methods is advantageous in terms of interpretability of predictions—a highly important aspect in clinical domain—whereas it still represents a significant issue for more complex methods.

Our review has retrieved only a few studies on the topic of extracting word embeddings from clinical notes. This may be due to insufficient available data to train the algorithms as well as the fact that embedding methods have been developed only recently. The issue of insufficient training data could be addressed using transfer learning methods, while using precomputed embeddings for specific diseases or categories of diseases could be useful to effectively capture longitudinal relationships.

Our review has shown that SVM and naïve Bayes algorithms were most often used for machine learning–based tasks or in combination with rule-based methods. This may be due to the popularity of these algorithms as well as because naïve Bayes, being a relatively simple algorithm, requires relatively small amount of training data (in comparison with deep classifiers, for example). Although it is not feasible to directly compare algorithmic performance of the studies that we considered (due to both diversity of data and challenges addressed), we have noted that the most commonly reported performance measures were sensitivity (recall), positive predictive value (precision), and *F* score.

Finally, our review has reinforced the fact that availability of public datasets remains scarce. This outcome was largely expected given the sensitivity of clinical data in addition to all the legal and regulatory issues, including the Health Insurance Portability and Accountability Act and the Data Protection Directive (Directive 95/46/EC) of the European Law (superseded by the General Data Protection Regulation 2016/679). As a result, the studies reviewed in this paper typically came from research-based health care institutions with in-house NLP teams having access to clinical data. Therefore, the need remains for shared tasks such as i2b2 and access to data that would increase participation in clinical NLP and contribute to improvements of NLP methods and algorithms targeting clinical applications.

## Limitations

This review has examined the last 11 years of clinical IE applications literature and may have the following limitations. The review is limited to journal articles written in the English language, and papers written in other languages, especially papers that consider clinical narratives, may provide additional results. In addition, papers using clinical articles from non-EHR systems have not been considered. Finally, focusing on the clinical domain may have introduced a bias with respect to the methods reviewed (rule-based vs machine learning), as rule-based methods are more prevalent in the clinical domain compared with other domains [122].

## Recommendations

Our review has shown that there is a clear necessity for clinical NLP methods to evolve beyond extraction of clinical concepts and focus more on concept understanding (ie, not only understanding of relationships between concepts but incorporation of clinical facts, domain knowledge, and general knowledge in the reasoning process). In this review, we have not encountered work that attempts to bridge the gap between concept extraction and concept understanding.

We have devised the following specific recommendations:

1. Focus on recognition of relationships among clinical concepts and entities. While progress has been made in recognizing entities in textual narratives (such as diseases, drugs, procedures), further efforts must be focused on automatic inference of relationships between these entities (for example, drug A causes adverse event B for chronic disease C), which in turn would allow deeper understanding of clinical text.
2. Temporal extraction, automated mark-up and normalization of temporal information from natural language texts, is an important aspect. This is especially relevant for clinical text as disease progression and clinical events are typically recorded chronologically, with specific events being significant only in a particular temporal context. As such, significant attention should be given to temporal extraction considering its implication in clinical context, especially since none of the works in this review dealt with temporal extraction (or used crude methods such as timestamps of clinical notes).
3. Scarcity of annotated clinical corpora has raised the need to exploit alternative sources of domain knowledge. In addition to mainstream sources such as biomedical literature, encyclopedias, and textbooks, automatic diagnostic and decision support systems could be exploitable (such as DXplain [123]). Transfer learning, a method of transferring knowledge from existing corpora in other domains to the clinical domain, also holds great potential and should be investigated in more detail.
4. Significant advances in effective clinical NLP will depend on large-scale corpora becoming available to researchers. While shared tasks such as i2b2 and its successor n2c2 are steps in the right direction, further incentives will be required such as developing mechanisms that would empower patients to donate their anonymized data or even providing algorithms that run on clinical text inside care institutions.

## Acknowledgments

This work was partially supported by the European Union's Horizon 2020 research and innovation program under grant agreement #769765.

## Conflicts of Interest

None declared.

## Multimedia Appendix 1

Search strategy.

[[PDF File (Adobe PDF File), 178KB](#) - medinform_v7i2e12239_app1.pdf ]

## Multimedia Appendix 2

Complete list of reviewed papers, chronic diseases and their classifications, algorithms used, publication venues, and excluded papers.

[XLSX File (Microsoft Excel File), 107KB - medinform_v7i2e12239_app2.xlsx ]

**Abbreviations**

    **BI-RADS:** Breast Imaging-Reporting and Data System
    **CHF:** congestive heart failure
    **CRF:** conditional random field
    **DeepPhe:** Cancer Deep Phenotype Extraction
    **EHR:** electronic health record
    **EMR:** electronic medical record
    **HF:** heart failure
    **i2b2:** Informatics for Integrating Biology and the Bedside
    **ICD:** International Classification of Diseases, 10th Revision
    **IE:** information extraction
    **MIMIC II:** Multiparameter Intelligent Monitoring in Intensive Care II
    **NLP:** natural language processing
    **PAD:** peripheral arterial disease
    **PRISMA:** Preferred Reporting Items for Systematic Reviews and Meta-Analyses
    **SVM:** support vector machine
    **THYME:** Temporal Histories of Your Medical Event